\documentclass[tightenlines,
twocolumn,prl,notitlepage,nofootinbib,preprintnumbers,longbibliography]{revtex4-1}

\usepackage{graphicx}
\usepackage{amsmath}
\usepackage{amsfonts}
\usepackage{amssymb}
\usepackage{float}
\usepackage[colorlinks]{hyperref}
\usepackage{multirow}
\usepackage{bbold}
\usepackage{soul}
\usepackage{makecell}
\usepackage{minibox}
\usepackage{soul}
\usepackage{fancyhdr}

\usepackage{mathtools}

\usepackage{amsthm}

\begin{document}

\preprint{LLNL-JRNL-823255}
\title{Scalability of gadolinium-doped-water Cherenkov detectors\\ for nuclear nonproliferation}%

\author{Viacheslav~A.~Li}
\email{corresponding author: li68@llnl.gov}
\affiliation{Lawrence Livermore National Laboratory, Livermore, CA 94550}
\author{Steven~A.~Dazeley} 
\affiliation{Lawrence Livermore National Laboratory, Livermore, CA 94550}
\author{Marc~Bergevin}
\affiliation{Lawrence Livermore National Laboratory, Livermore, CA 94550}
\author{Adam~Bernstein}
\affiliation{Lawrence Livermore National Laboratory, Livermore, CA 94550}

\date{April 18, 2022}

\begin{abstract}
   Antineutrinos are an unavoidable byproduct of the fission process. The kiloton-scale KamLAND experiment has demonstrated the capability to detect reactor antineutrinos at few-hundred-km range. But to detect or rule out the existence of a single small reactor over many km requires a large detector. So large in fact that the optical opacity of the detection medium itself becomes an important factor. If the detector is so large that photons cannot traverse across the detector medium to an optical detector, then it becomes impractical. For this reason, gadolinium-doped-water Cherenkov detectors have been proposed for large volumes, due to their appealing light-attenuation properties. Even though Cherenkov emission does not produce many photons and the energy resolution is poor, there may be a place for Gd-doped-water detectors in the far-field nuclear reactor monitoring.
   In this paper, we focus on the reactor discovery potential of large-volume Gd-doped-water Cherenkov detectors 
   for nuclear nonproliferation applications. Realistic background models for the worldwide reactor flux, geo-neutrinos, cosmogenic fast neutrons, and detector-associated backgrounds are included. We calculate the detector run time 
   required to detect a small 50-MWt reactor at a variety of stand-off distances as a function of detector size.  We highlight that at present, PMT dark rate and event reconstruction algorithms are the limiting factors to extending beyond $\sim$50-kt fiducial mass.
\end{abstract}
\maketitle

\section{Introduction}
Nuclear reactors produce enormous numbers of antineutrinos as a consequence of the production of beta-decaying neutron-heavy fission daughters. In fact, a typical large nuclear power reactor ($\sim$1 GWt) generates more than $\sim$10$^{19}$ fissions per second and $~\sim$10$^{20}$ antineutrinos per second during normal operation. %
Each fissioning isotope produces its own spectrum of daughter isotopes. Therefore, the resulting antineutrino flux and energy spectrum is sensitive to the fuel content~\cite{Borovoi1978}. %
Furthermore, since antineutrinos only interact weakly with matter, they are impossible to shield~\cite{Fermi:1934hr}. Therefore, information about the operational status and fuel content of a reactor can in principle be monitored remotely at significant distances via the antineutrino flux. 

Although growth in the worldwide number of reactors since the mid 1980s has been slow, about 500 nuclear power reactors world-wide currently account for $\sim$10\% of global electricity production. 
Despite being a carbon-free energy source, one of the reasons for the slow nuclear growth rate is the enormous capital cost of building a new nuclear reactor. Future reactor designs address this issue by attempting to go smaller and simpler. Small Modular Reactors (SMRs), are designed to be simpler to operate and reduce capital costs~\cite{Bays_SMR_INMM2021}. Some new reactor designs utilize smaller fuel elements, or liquid fuel. These designs may be more difficult to safeguard using present-day techniques which emphasize item accountancy. 

Recently, methods for monitoring the operational status of future reactor designs have been proposed that don't rely on item accountancy. One such method is to monitor the antineutrino emissions of these reactors. The reality is however, that antineutrinos are extremely difficult to detect for the same reasons they are impossible to shield~\cite{Cowan1956}. The range over which it is practical to monitor a reactor will be limited by cost considerations related to detector size and the requirement to place such detectors underground to shield from cosmic-ray-induced backgrounds. 

In recent years, reactor antineutrino detectors have been deployed at 1--2 km distances to measure neutrino-oscillation properties~\cite{Boehm:1999gk, Apollonio:2002gd, Abe:2011fz, An:2012eh, Ahn:2012nd}, and a 1-kiloton detector called KamLAND, was situated at a flux weighted average distance of $\sim$180\;km from Japan’s reactor fleet to study oscillation at long range~\cite{Araki:2004mb}. In the very near future, the JUNO detector will commence operations at a distance of 53~km from two reactor complexes in China~\cite{An:2015jdp}; while the SNO+ detector --- in Canada, will be able to observe several CANDU reactors with an effective baseline of a couple of hundred kilometers~\cite{Baldoncini:2016mbp}. All of these experiments use liquid organic scintillator as the detection medium. The fundamental limit to the size of these detectors is the attenuation of light, which is $\sim$10--20\;m in liquid scintillator.

Gd-doped water is another medium for reactor-antineutrino detection, which makes use of Cherenkov light instead of scintillation light.
One existing example is  Super-Kamiokande~\cite{Super-Kamiokande:2021the}, which %
has completed an initial fill of Gd-H$_2$O using gadolinium sulfate octahydrate at approximately one tenth of the goal concentration of 0.2\% by mass. 
While Cherenkov emission produces less light with poorer energy resolution, attenuation lengths approaching $\sim$100\;m are achievable with Gd-doped water, which might enable very large detectors~\cite{Ikeda:2019pcm}. 

In 2015, the WATCHMAN collaboration proposed to demonstrate monitoring of a single reactor site using the Hartlepool reactor, with a detector to be deployed 26 kilometers away in the Boulby mine~\cite{Askins:2015bmb}. Planning for this effort ceased following a UK government announcement in 2021 of premature shutdowns of the Hartlepool and other reactors, which would have compromised the deployment schedule for the Boulby site.

Gadolinium-doped water is sensitive to electron antineutrinos via the inverse beta decay (IBD) reaction:

\begin{equation}
    \bar\nu_e + {}^{1}\mathrm{H} \to e^+ + n
\end{equation}

For reactor antineutrinos, the prompt ($e^{+}$) is ~few MeV. The Cherenkov light from the positron (few MeV) is detectable, however the 511-keV gammas do not deposit a sufficient energy to generate Cherenkov light
~\cite{Cherenkov:1934zz,Frank:1937fk}.
The neutron captures primarily on $^{155}$Gd or $^{157}$Gd, producing a few gamma rays of few-MeV each~\cite{Hagiwara:2018kmr,Tanaka:2019hti}, which sum to ~8 MeV. 
The gammas then Compton-scatter, producing sufficient Cherenkov light to permit detection of a delayed event. The average delay between the positron flash and the thermal-neutron capture depends on the concentration of gadolinium in the water. At $0.1\%$, the neutron capture time is $\sim 30 \; \mu$s.

To collect the Cherenkov photons, the detector requires a sufficiently high coverage of photosensors. In the Super-Kamiokande, about 40\% of the detector wall area is effectively covered with photomultipliers tubes. Light attenuation is on the order of a few hundred meters~\cite{Pope:97, Abe:2013gga} at the blue wavelength suitable for PMT detection in large detector volumes.
At few MeV energies, only a handful of photo electrons are produced. So, in addition to high photocathode coverage of the detector, a high quantum efficiency, low radioactivity, and low dark noise are desired characteristics of the PMTs.

For nonproliferation and science alike, 
Gd-doped water may be a cost-efficient and environmentally-friendly alternative compared to liquid-scintillator based detectors.
In the early 2000s, there were several proposals to dope water Cherenkov detectors with gadolinium in order to be sensitive to the inverse-beta-decay reaction~\cite{Bernstein2001, beacom_vagins_2004}. 
Since then many projects have been pursued to study the feasibility of the technique, summarized in Table~\ref{tab_Gd_projects}.
Measurements of the effect of Gd doping on water transparency have been performed with EGADS~\cite{Ikeda:2019pcm}, the largest engineering/physics demonstration to date. 

\begin{table}[]
    \centering
    \small
    \begin{tabular}{|l|c|c|r|}\hline
        Project & Mass [tons] & Gd content & Ref.  \\\hline\hline
Watanabe et al. & 0.002 & 0.2\% wt. GdCl$_3$ & \cite{WATANABE2009320}\\\hline %
WAND & 1.0 & 0.4\% wt. GdCl$_3$ & \cite{DAZELEY201532}\\\hline
ANGRA & 1.3 & 0.2\% wt. GdCl$_3$ & \cite{angra2019}\\\hline
WATCHBOY & 2 & 0.2\% wt. GdCl$_3$ & \cite{Dazeley:2015uyd}\\\hline
ANNIE & 26 & 0.2\% wt. Gd$_2$(SO$_4$)$_3$ & \cite{Back:2017kfo}\\\hline
EGADS & 200 & 0.2\% wt. Gd$_2$(SO$_4$)$_3$ & \cite{Ikeda:2019pcm} \\\hline
WATCHMAN  & 6,000 [1,000]  & 0.2\% wt. Gd$_2$(SO$_4$)$_3$ & \cite{Askins:2015bmb} \\\hline
SuperK-Gd & 50e3 [22.5e3] &  0.2\% wt. Gd$_2$(SO$_4$)$_3$ & \cite{Super-Kamiokande:2021the} \\\hline
\end{tabular}
    \caption{Global research efforts on Gd-H$_2$O technology. Gadolinium is added in the form of either gadolinium chloride or sulfate, and has natural isotopic abundances (no isotopic enrichment). For WATCHMAN and SuperK, a corresponding fiducial mass is also listed in square parentheses.}
    \label{tab_Gd_projects}
\end{table}

\begin{figure}[ht]
\centering\includegraphics[width=1.\linewidth]{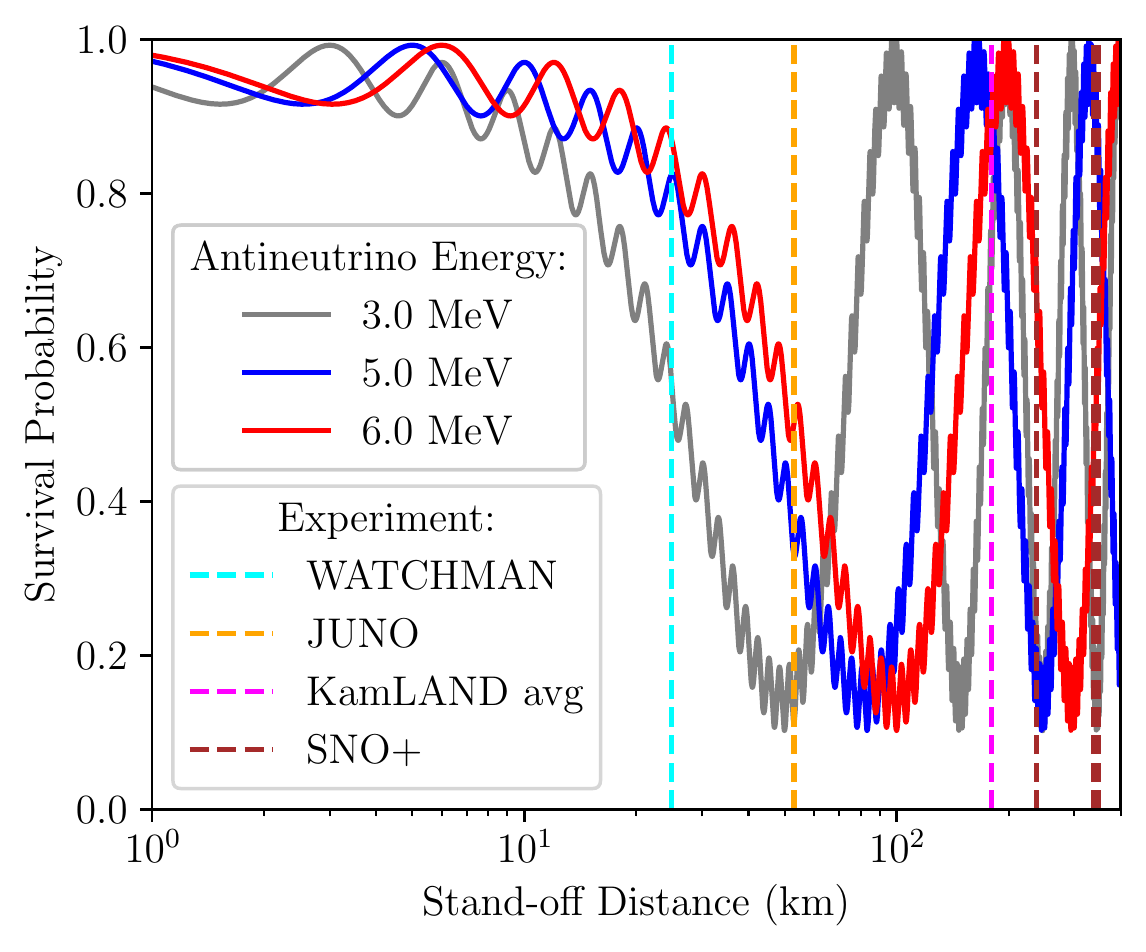}
\caption{Survival probability of electron antineutrinos as a function of distance, with a number of far-field detector baselines shown (for KamLAND, a flux-weighted average baseline is plotted). Antineutrinos with energies higher than about 4 MeV become visible in a water-Cherenkov detector. These are significantly suppressed at $\sim$50--100-km distances (the ``death valley'' for reactor antineutrinos).}
\label{fig_osc}
\end{figure}

The number of IBD interactions inside a detector located at some distance from a reactor source can be calculated as follows. 
Each fission results in $\sim$6 antineutrinos on average.
Approximately a quarter of these are 
above the IBD 1.8-MeV threshold. 
Since the antineutrinos are emitted isotropically, %
the IBD interaction rate can be estimated as:
\begin{equation}
    \begin{split}
       N_{IBD} \cong 1.5 P_{surv} \times
    \left( \frac{power}{50\ \mathrm{MWth}} \right) \times
    \left( \frac{mass}{10\ \mathrm{kt}} \right) \times\\
    \times \left( \frac{time}{1\ \mathrm{day}} \right) \times
    \left( \frac{distance}{ 10\ \mathrm{km}} \right)^{-2}
    \end{split}
\end{equation}
where $P_{surv}$ is a probability of electron antineutrino to retain its flavor --- the so-called survival probability. It is important to note that the survival probability is a function of antineutrino energy and distance. 
Survival probability factor $P_{surv}$, due to neutrino oscillations, can contribute up to a factor of few suppression, depending on the energy and stand-off distance.
For example, the effect of oscillations is especially pronounced for 5-MeV antineutrino at $\sim$100-km distance, as shown in Fig.~\ref{fig_osc}.

\begin{figure}[ht]
\centering\includegraphics[width=1.\linewidth]{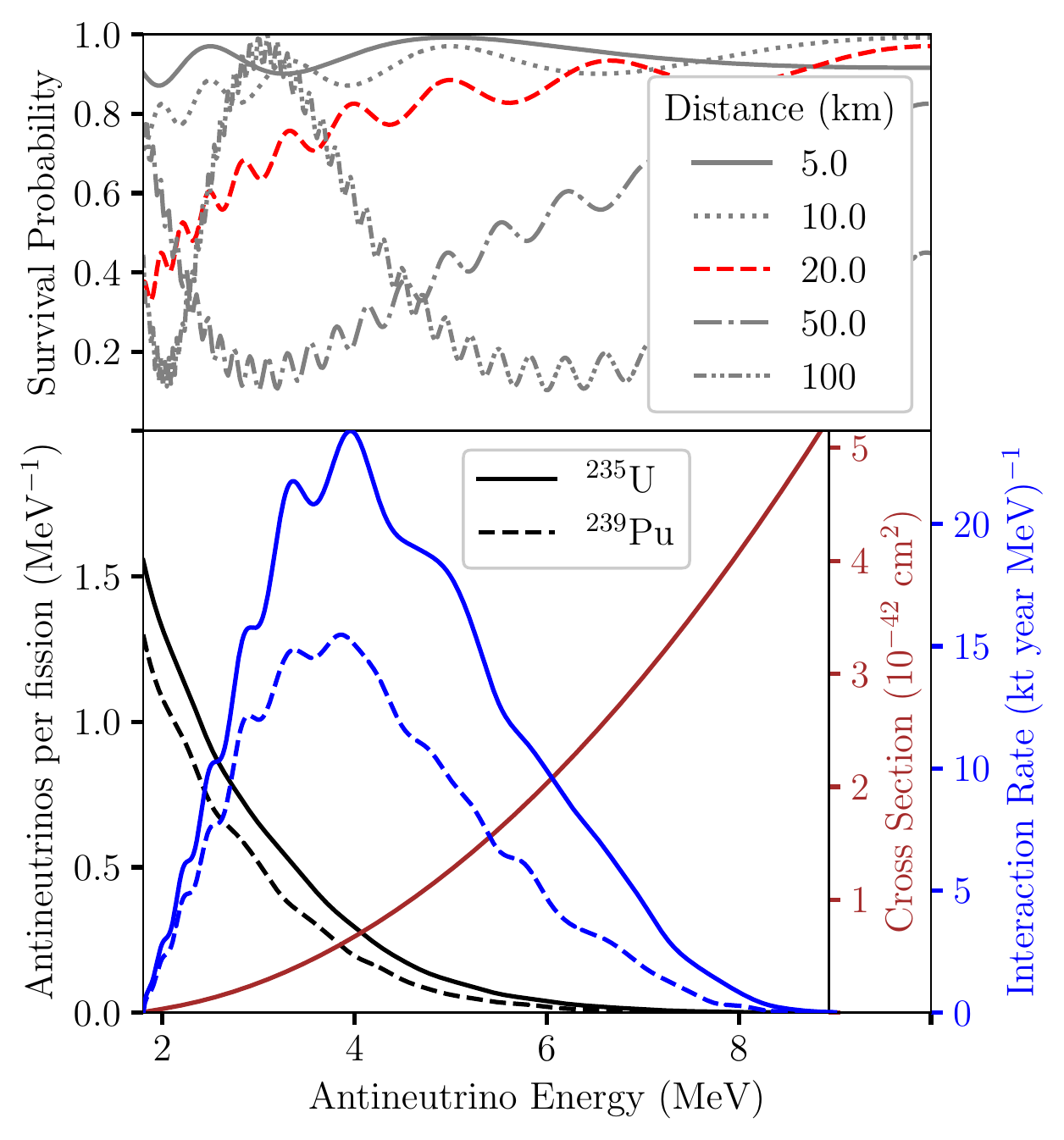}
\caption{An antineutrino spectrum at the detector location would have a form shown in blue lines, which takes into account an emitted spectrum, fission energy per isotope, antineutrino survival probability, and the IBD cross-section. The full Strumia-Vissani approximation for the IBD cross-section~\cite{Strumia:2003zx} and Huber model for the emitted reactor spectrum~\cite{Huber:2011wv} are shown. An IBD interaction rate in a kiloton water detector from 50-MWt reactor at 20-km stand-off distance with a two hypothetical cores --- 100\%  $^{239}$Pu fission fractions (dashed blue line) and  100\% $^{235}$U fission fractions (solid blue line); a realistic core would lie in between these two curves. {\it Top panel:} survival probability of reactor antineutrinos as a function of antineutrino energy for selected stand-off distances; the 20-km curve, highlighted in red, is folded to get the interaction rate shown in the bottom plot (blue curves).}
\label{fig_nuebar_spectrum}
\end{figure}

\begin{figure}[ht]
\centering\includegraphics[width=1.\linewidth]{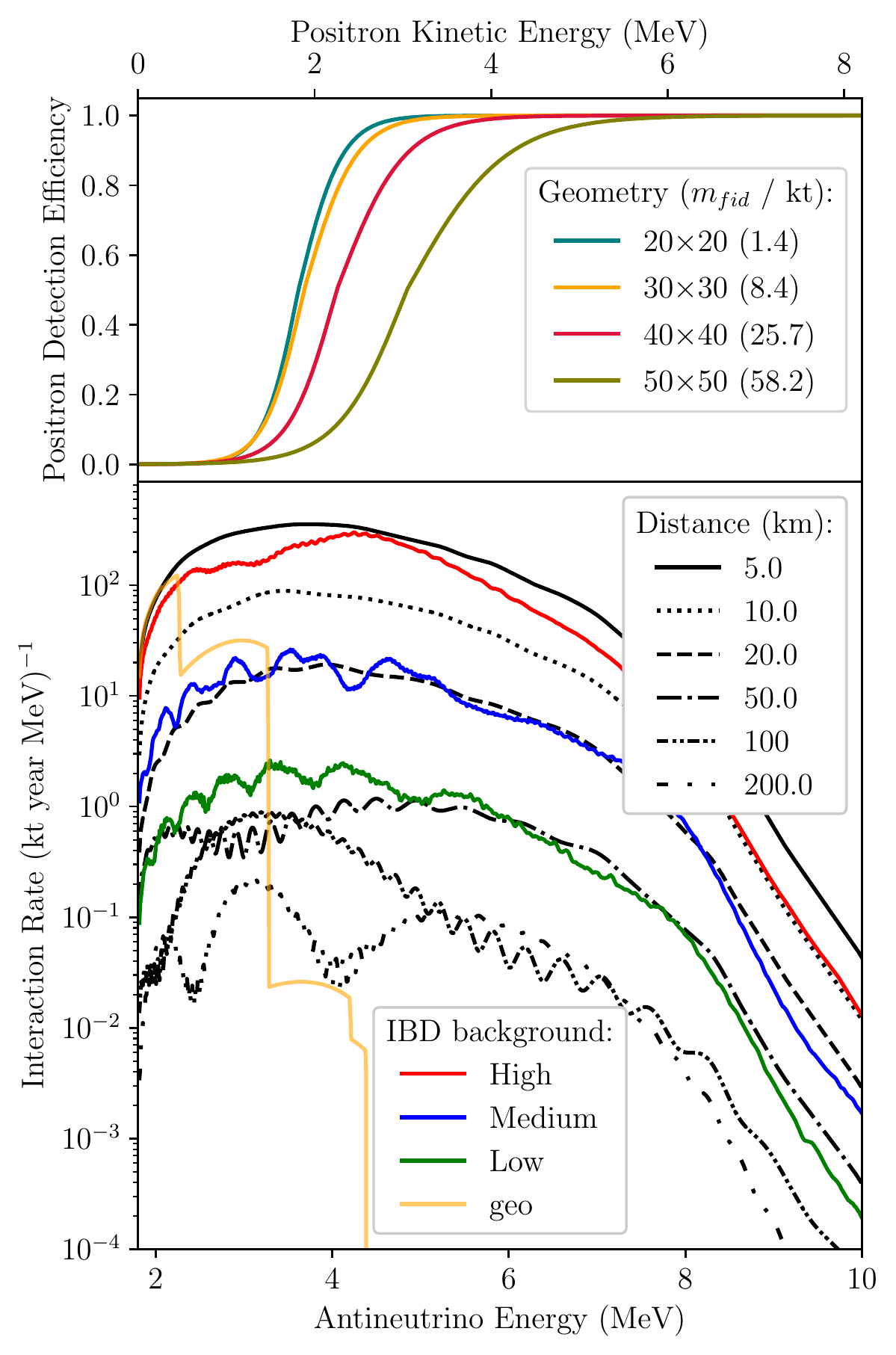}
\caption{IBD interaction rate as a function of antineutrino energy for three locations. For reactor the power is set at 50~MWt and the fuel composition is PWR, using Huber model for the emitted reactor spectrum. Each of the positron detection efficiency curves shown (on the top plot) were generated assuming a minimum number of PMT hits. The PMT hit threshold can be optimized for different detector sizes. The details of this optimization are presented in the next section. The geo-neutrinos are invisible in water-Cherenkov detectors (for geo-neutrinos a location at Frejus is shown as an example).}
\label{fig_nuebar_spectrum_3locations}
\end{figure}

As shown by~\cite{Bernstein:2019hix}, for the purposes of nuclear nonproliferation, even a relatively small reactor of 50~MW thermal power is capable of producing a significant quantity of plutonium (8 kg) in one year.
With the assumptions of 100\% detection efficiency and zero detector background, the study provided estimates of the time required to detect such a reactor under various world-reactor-background conditions. 
By contrast, this paper takes into account more realistic efficiency estimates and detector related backgrounds; thus, the dwell times obtained are longer, but might be considered more realistic. %
To account for the sometimes large variations in the world-wide reactor-antineutrino background, the same three test locations are assumed here: Andes, Baksan, and Frejus. The closest power reactor at the Andes location is 563~km away, at Baksan --- it is 360-km distant, and at Frejus --- 133~km.
The world-reactor background varies by approximately a factor of 100 between Low, Medium, and High test locations, as listed in Table~\ref{tab_3loc}. %

Fig.~\ref{fig_nuebar_spectrum} shows the antineutrino flux spectrum, the IBD interaction cross section, and the IBD interaction spectrum that can be expected from a hypothetical 50-MWt test reactor located at a distance of 20~km from a hypothetical detector.
The antineutrino spectrum from our test reactor is calculated based on the reactor power, fuel content, and distance to the detector. %
Neutrino oscillations become a significant factor in the far-field. Fig.~\ref{fig_nuebar_spectrum_3locations} shows the antineutrino spectrum normalized by $1/L^2$, and folded with antineutrino survival probability. Note that the antineutrino fluxes for 100-km and 200-km baselines are comparable above 5-MeV antineutrino energy due primarily to the effect of oscillations. 
The low-energy antineutrinos are impacted by the detector threshold effects  in large water-Cherenkov detectors. This detector threshold effect is illustrated in Fig.~\ref{fig_nuebar_spectrum_3locations}, and will be further detailed in the next section.

To obtain IBD event rates in a realistic detector, we  implement an algorithm approximated by the  formula:

\begin{equation}
\begin{split}
        R_{d} =
    \left(
    \int 
    R_{e}
    \sigma_{IBD}
    P_{surv}
    \frac{1}{4\pi L^2}
    \epsilon_+
    d E_+
    \right)
    \epsilon_n
    N_t
    f_{2\mathrm{m}}
    f_{100\mu\mathrm{s}}
\end{split}
\label{eq_IBD_rates}
\end{equation}
where $R_{d}$ is the IBD detected rate; $R_{e}$ --- the emitted antineutrino spectrum (taking into account reactor power and fuel composition); 
$\sigma_{IBD}$ --- the IBD cross-section (a function of energy); 
$P_{surv}$ --- antineutrino survival probability (a function of energy and distance); 
$L$ --- distance between the reactor and the detector; $\epsilon_+$ --- positron detection efficiency (a function of energy); 
$E_+$ --- positron energy; $\epsilon_n$ --- neutron detection efficiency; 
$N_t$ --- number of targets (hydrogen nuclei) in the fiducial volume; 
$f_{2\mathrm{m}}$ --- 2-meter spacial proximity coefficient (in the range of 91\%--98\%); and 
$f_{100\mu\mathrm{s}}$ --- 100-microsecond temporal proximity coefficient (set at 98\%). The fuel composition is kept constant in this study, and is listed in Table~\ref{tab_reactor_fuel}.

\begin{table}[t]
\small
\centering
\begin{tabular}{|l||c|c|c|}
\hline
Background & Total rate & CR rate  & CR distance  \\\hline \hline
Low  & 6.6    &  1.6  &       563 km \\\hline
Medium  & 65.4   &  2.5  & 	  360 km \\\hline
High & 798.7  &  64.3 & 	  133 km \\\hline
\end{tabular}
\caption{Total world-reactor background rates, in units of IBD interactions per kiloton of water per year, for three representative locations that have a High, Medium, and Low reactor-antineutrino flux. Closest reactor (CR) rate and distance are also listed for each location.}
  \label{tab_3loc}
\end{table}

\begin{table}[t]
\small
\centering
\begin{tabular}{|l||c|c|c|c|}
\hline
 & $^{235}$U &  $^{238}$U & $^{239}$Pu & $^{241}$Pu\\\hline \hline
Fission fraction  &  0.56 & 0.08 & 0.30 & 0.06 \\\hline
Energy / fission [MeV] &   201.9 & 205.0  & 210.9 & 213.4 \\\hline
\end{tabular}
\caption{Fuel composition (fission fractions) and energy release per fission used in this study. }%
  \label{tab_reactor_fuel}
\end{table}

\section{Detector Description}

In the following, we investigate the sensitivity of large water Cherenkov detectors doped with gadolinium. We are interested in estimating the detector size required to successfully detect within one year a test reactor at various distances. In all cases, the assumed detector is a right cylinder shape (like Super-K), with photocoverage is fixed at 40\%.
The veto thickness (to suppress cosmogenic spallation neutrons from the surrounding rock) is fixed at 2 meters. The inner detector is also a right cylinder. The PMTs are modeled on 10-inch Hamamatsu 7081 PMTs equiped with low-activity glass. An average dark rate of 3~kHz per PMT is also assumed, based on internal measurements of a selection of about a hundred PMTs. All inward facing PMTs are placed perpendicular to the wall.
The fiducial volume boundary is placed at 2 meters from the PMT wall to reduce PMT-based radiation background. 

Event position reconstruction is done using a software package adapted from Super-K called BONSAI~\cite{Smy:2007maa}. For a given event energy, the detector response depends somewhat on the event position. However for small detectors, the detector response over the fiducial volume is approximately constant. For large detectors, light attenuation can become significant, so the response of events in the center of the detector is generally slightly smaller than the response near the edge of the fiducial volume. Nevertheless, for this work we have used the approximation that the number of detected photons detected within a Cherenkov wavefront is proportional to the energy of a (minimum ionizing) particle. To count Cherenkov photons, the time residuals of the position reconstructions are used: n9 is the number of PMTs with time residuals between --3\;ns and +6\;ns (values are chosen based on PMT timing uncertainty and expected light scattering). In the following, n9 will serve as a rough energy estimator. The ``n9'' parameter is modeled after a similar parameter used at Super-K~\cite{SuperK_DAQ, Abe:2016nxk}, using a larger time range (18 nanoseconds). The smaller time window was considered appropriate for this study due to the improved time resolution of the Hamamatsu 10" PMTs.
A set of detector geometries of between 0.3 kt and 58 kt fiducial were studied here, and listed in Table~\ref{tab_geom}.

\begin{table}[ht]
    \centering
    \begin{tabular}{|l|r|r|r|r|}\hline
d [m] $\times$ h [m] & $m$ [kt] & $m_f$ [kt] & \# of 10" PMTs  \\\hline\hline
15$\times$15	& 2.7	& 0.3  & 4512   	\\\hline
20$\times$20	& 6.2	&  1.4 & 9516   	\\\hline
30$\times$30	& 21.2	& 8.4 & 25182   	\\\hline
40$\times$40 	& 50.3 & 25.7 & 48258  	\\\hline
50$\times$50	& 98.2	& 58.2 & 78856  	\\\hline

   \end{tabular}
    \caption{Simulated detector geometries. The fiducial volume is defines as 4 meters from the outer tank, or 2 meters from the PMTs. Other geometries are being interpolated based on the simulated ones.}
       \label{tab_geom}
\end{table}

\begin{figure}[ht]
\centering
\includegraphics[width=.7\linewidth]{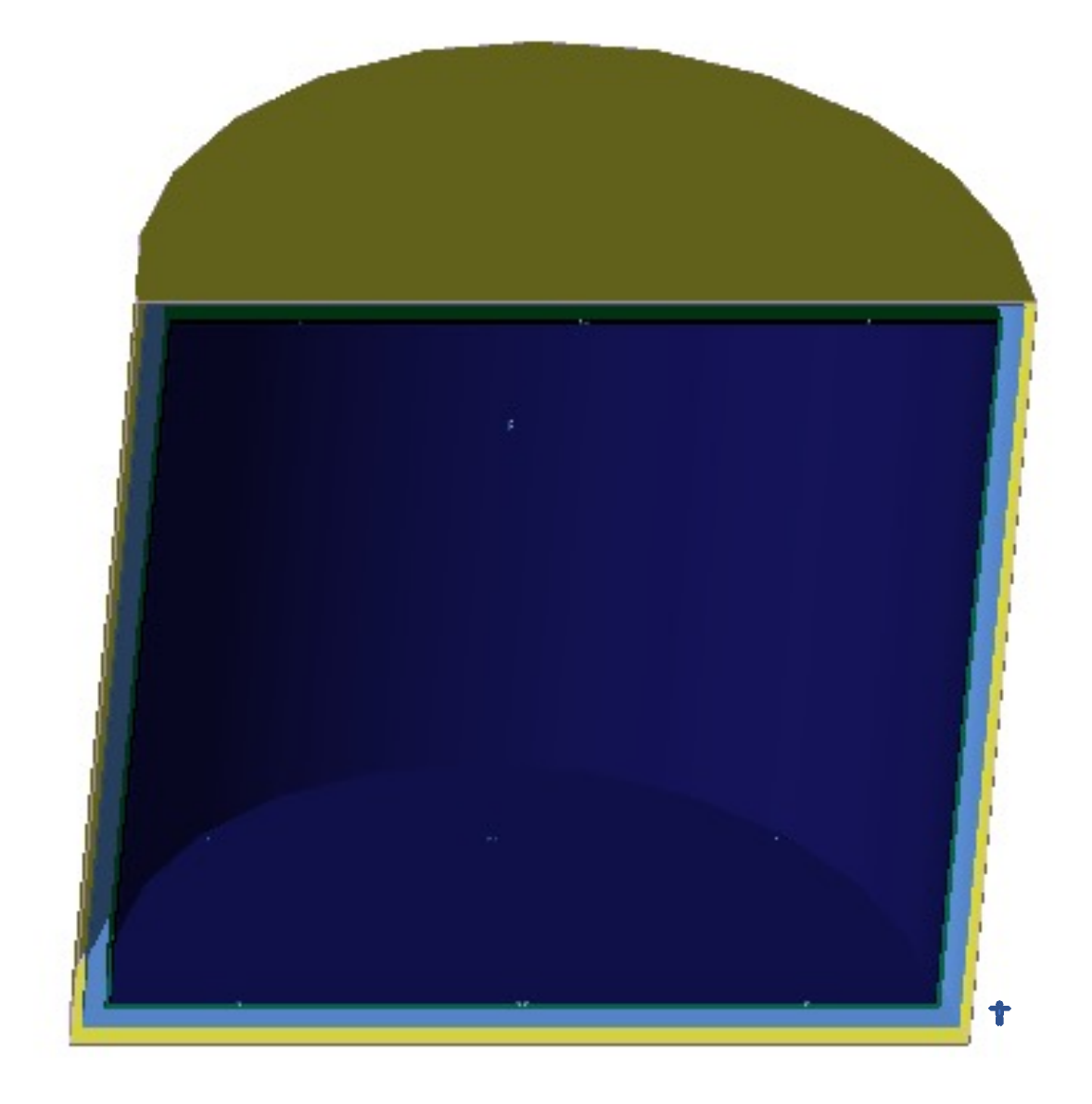}
\caption{GEANT4 (RAT-PAC) visualization of one of the detector geometries. The distance (outer veto region) between the inner tank and the outer tank is kept at 2 meters in all simulations. In this particular view, a cutaway of the 50$\times$50 geometry is presented with only a few 10-inch PMTs (those little dots on the image) to demonstrate the scale. A figure of a person is shown approximately to-scale.}
\label{fig_geom_G4}
\end{figure}

A GEANT4-based simulation package called RAT-PAC~\cite{rat} is used throughout this study. While based on GEANT4~\cite{geant4}, it includes
useful features specifically designed for reactor-antineutrino detector simulations. It was recently further adapted by the WATCHMAN collaboration ~\cite{Askins:2015bmb}. 
A visualization of a 50$\times$50 geometry is shown in Fig.~\ref{fig_geom_G4}.

\begin{figure}[tbp]
	\begin{center}
\includegraphics[width=1.\linewidth]{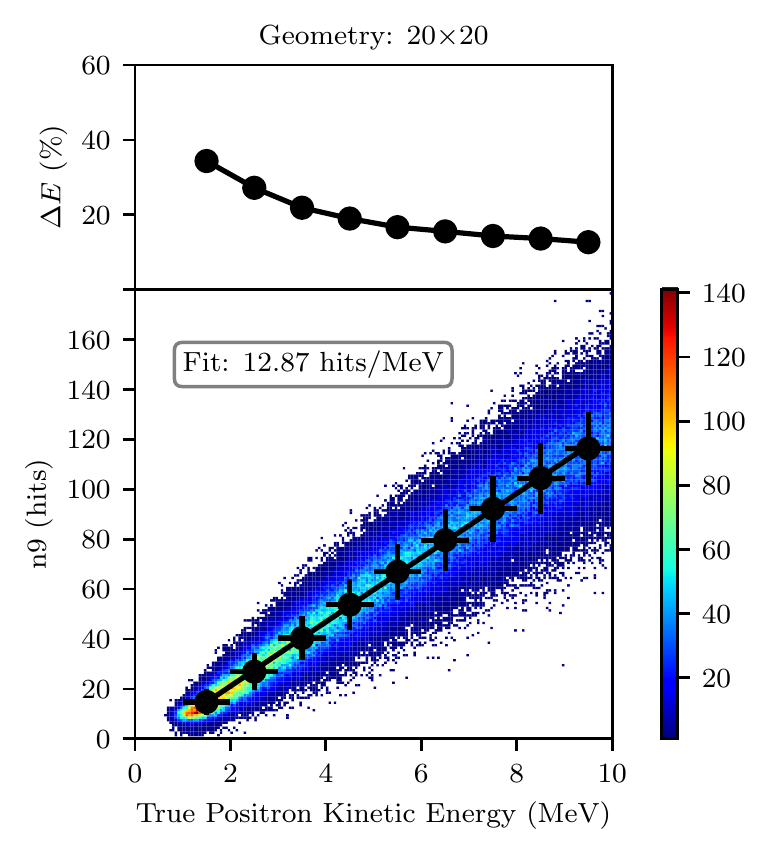}
\includegraphics[width=1.\linewidth]{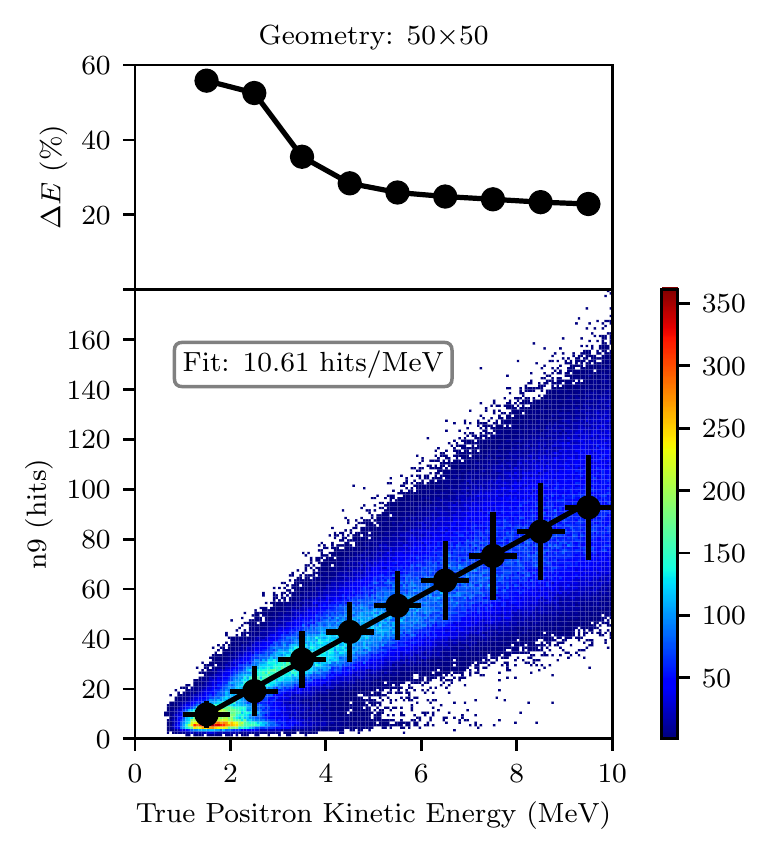}
\end{center}
\caption{The reconstructed number of PMT hits within 9-nanosecond window (n9 parameter) versus positron kinetic energy (Monte Carlo true) for a flat [0, 10]-MeV positron spectrum 
in the 
fiducial volume. 
A detector trigger threshold of 10 PMT hits was applied. {\it Top panel:} 20$\times$20 geometry; {\it bottom:} 50$\times$50. The comparison shows that a degradation in energy resolution $\Delta E$ accompanies increased detector sizes. 
The large number of events in the bottom left corner of the plot for the 50$\times$50 geometry indicates  mis-reconstruction of low-energy events.} %
\label{fig_flat_positron_n9}
\end{figure}

Physics events in the few MeV region only produce a few dozen photoelectrons. Most PMTs register only one photoelectron. %
Fig.~\ref{fig_flat_positron_n9} shows the relation between the n9 parameter and the true positron energy, as simulated in RAT-PAC, along with a plot of energy resolution versus the energy.

\begin{table}[h]
    \centering
    \begin{tabular}{|l||r|r|r||r|r|}\hline
Geometry & n9$_{pr}$ & n9$_{del}$ & 2-m fraction & $f_\mathrm{2m}$ & $f_\mathrm{100\mu s}$ \\\hline\hline
15$\times$15	& 17	&  25 & 0.05256  &
99\%	& 98\%\\\hline
20$\times$20	& 17	&  25 & 0.02469 &
99\%	& 98\%\\\hline
30$\times$30	& 18	& 25 & 0.00401  &
97\%	& 98\% \\\hline
40$\times$40 	& 20 & 25 & 0.00130  & 
94\% & 98\% \\\hline
50$\times$50	& 26	& 26 & 0.00058  & 
91\%	& 98\% \\\hline
   \end{tabular}
\caption{Prompt n9 cuts obtained using optimization for four different detector sizes, as well as 2-meter bubble fractions used in estimation of IBD-like accidental backgrounds. The IBD spatial and temporal proximity coefficients are also listed, as used in IBD signal and background calculations.}
\label{tab_geom_optimal_param}
\end{table}

\begin{figure}[ht]
\centering
\includegraphics[width=1.\linewidth]{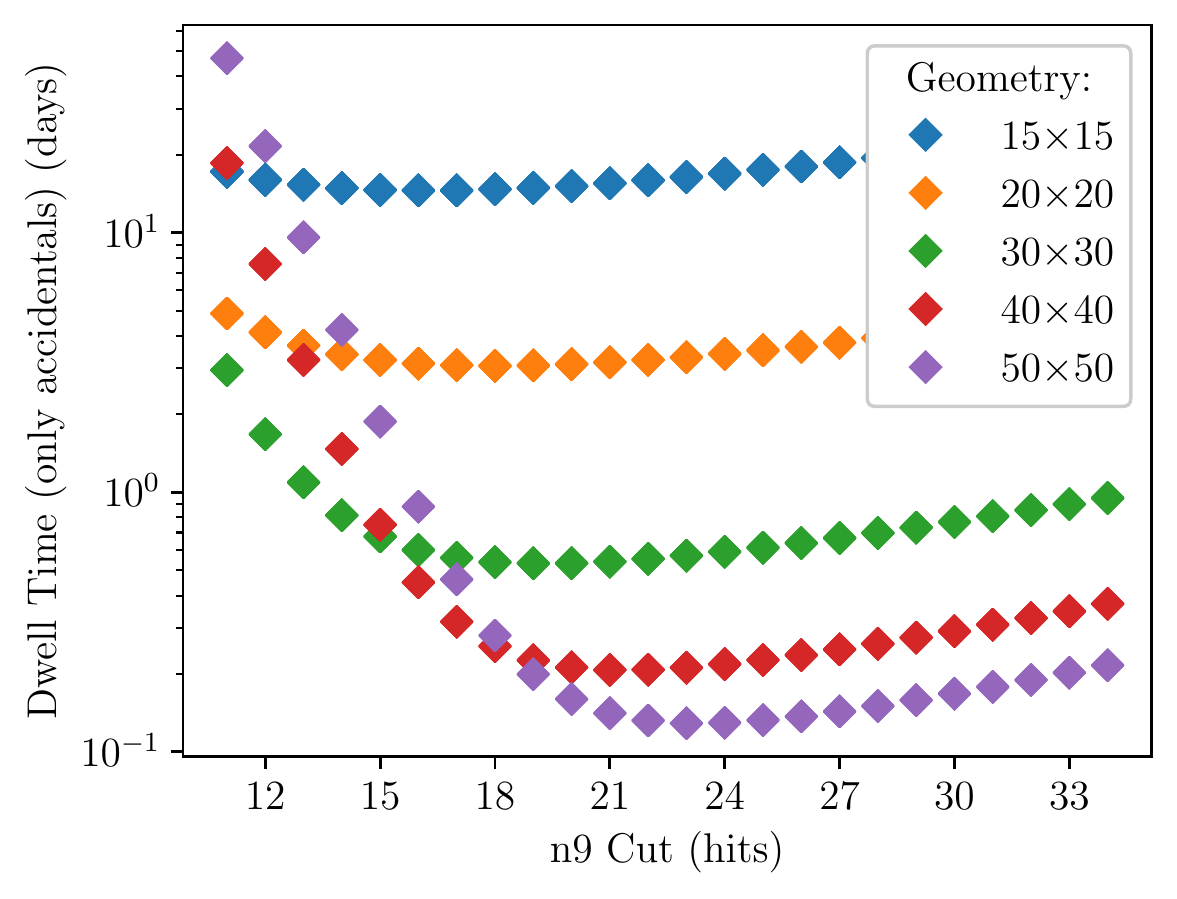}
\caption{The dwell-time metric based on Eq.~\ref{eq_currie_sys} including only detector accidental backgrounds to estimate optimal n9 cut on the prompt (minimum of these curves). For the 50$\times$50 geometry, an optimal n9 cut was chosen 26 to eliminate the effect of the PMT dark rate.}
\label{fig_accDwellTime}
\end{figure}

IBD event candidates were selected on the basis of characteristics: 1) prompt signal; 2) delayed signal; 3) the position in the detector; 4) the time and distance between correlated pairs. A summary of the analysis cuts used are presented in Table~\ref{tab_geom_optimal_param}. In order to simplify the analysis, and aid in scaling sensitivity between each of the simulated detector sizes, it was decided to fix some of the analysis cuts 
{\it a priori}. For example, the fiducial volume was defined to extend to 4~m from the tank wall for all detector sizes. The maximum distance and time between correlated event pairs were defined to be 2~m and 100~$\mu$s respectively. These initial guesses were informed by an initial round of simulations which indicated limited sensitivity to detector size. The remaining analysis cuts, the n9 prompt and delayed, are somewhat dependent on the detector size. However, in the interests of simplicity, since the n9 prompt cut was found to be the most sensitive to detector performance, we opted to optimize only n9 prompt, while setting the n9 delayed to a reasonable constant value (where possible). 
An optimization of the n9 prompt analysis cut was performed for each detector size (using detector dwell time as the metric), maintaining all the other analysis cuts constant, shown in Fig.~\ref{fig_accDwellTime}.
The primary detector background that impacts the n9 prompt cut optimization originates from radioactive decays in the PMT glass and the detection medium.
The central regions of the detector medium can be expected to be relatively radio-pure~\cite{Nakano:2019bnr}. 
Our simulations assumed levels of trace radioactive nuclei in the water consistent with Super-K. 
This resulted in a low level (but still significant) background event rate. The primary source of background events was due to the $^{238}$U and $^{232}$Th decay chains. The most problematic isotopes were $^{214}$Bi, $^{212}$Bi, and $^{208}$Tl.

\subsection{Detection efficiency}
\begin{figure}[ht]
\centering\includegraphics[width=1.\linewidth]{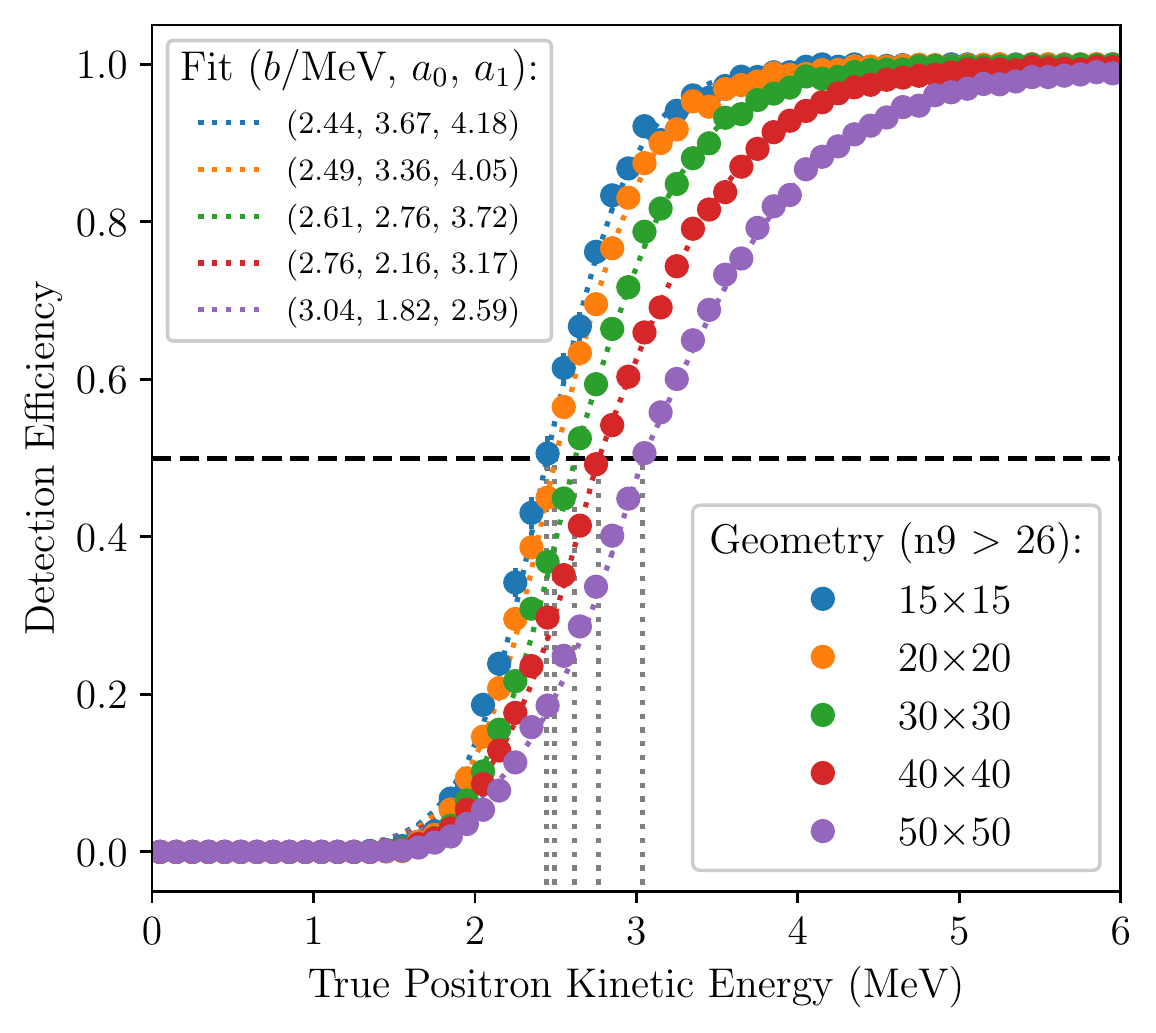}
\caption{Positron detection efficiency as a function of Monte Carlo positron kinetic energy. 
Efficiency curves for $n9 > 26$ are shown as an example here.
The vertical dashed lines correspond to the energies at 50\% efficiency level --- the {\it b} parameter in the fit.}%
\label{fig_det_eff}
\end{figure}

As mentioned previously, five distinct detector sizes were simulated. To determine the performance of intermediate detector sizes, a series of interpolations were used. The first step in this process was to generate a set of efficiency curves for positrons and neutrons.   
For each configuration in Table~\ref{tab_geom}, positrons were simulated with a flat spectrum [0, 10]-MeV distributed uniformly in the detector fiducial volume.
A set of efficiency curves was then generated and parametrized using a combination of two Fermi functions: %
\begin{equation}
\label{eq_eff}
    \epsilon = 
    \frac{1}{2}
    \left[
    \frac{1+\mathrm{sign}(E-b)}{1+\exp(-a_0 (E-b))} 
    +
    \frac{1+\mathrm{sign}(-E+b)}{1+\exp(-a_1 (E-b))} 
    \right]
\end{equation}
where $E$ is the positron kinetic energy and ``sign'' is a sign function, such as $\mathrm{sign}(x) = \pm 1$ if $x \gtrless 0$.
For each n9-parameter cut and each {\it simulated} geometry, we obtained efficiency curves and a set of stored fit parameters $a_0, a_1,$ and $b$ in a lookup table.
The $b$ coefficient is the energy at 50\% efficiency. A set of positron efficiency curves %
are shown 
in Fig.~\ref{fig_det_eff} for an n9 cut of 26 PEs.
The positron efficiency curves shown in Fig.~\ref{fig_nuebar_spectrum_3locations} were generated using the optimized n9 thresholds presented in Table~\ref{tab_geom_optimal_param}. 

\begin{figure}[ht]
\centering\includegraphics[width=1.0\linewidth]{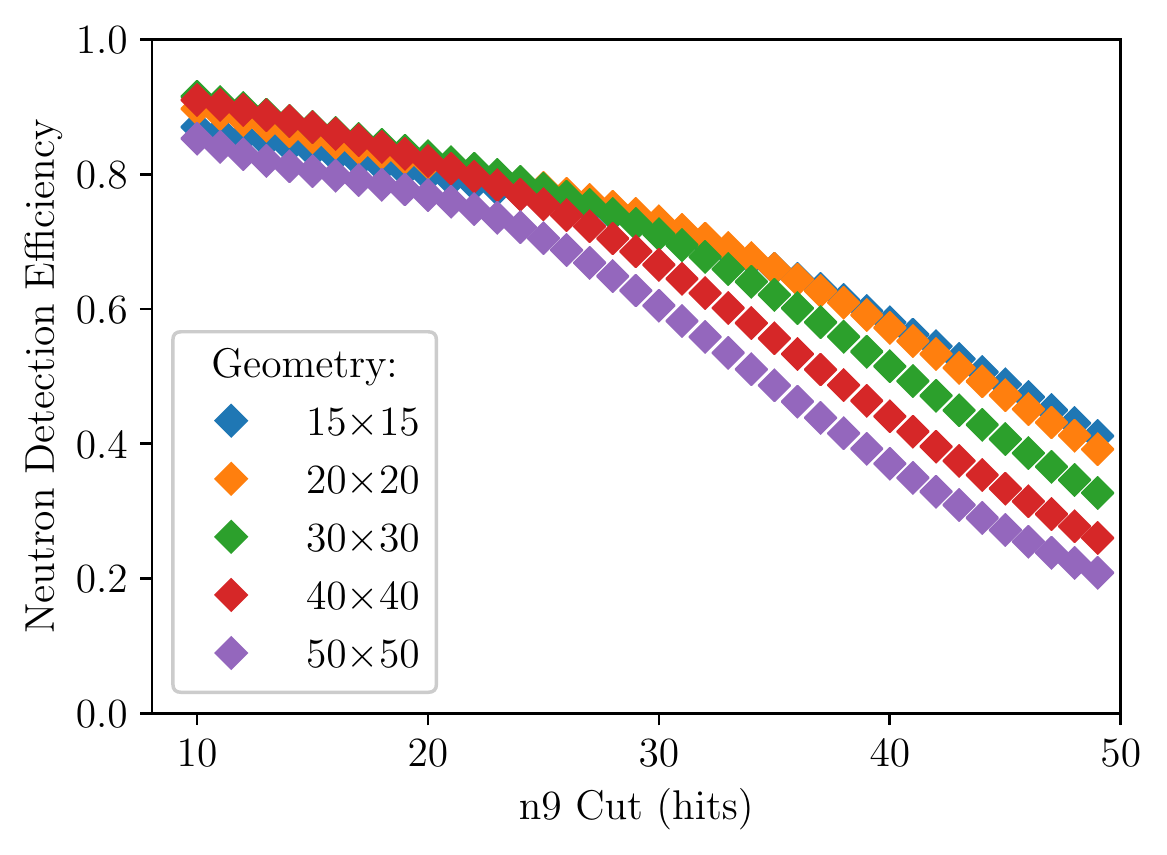}
\caption{Neutron detection efficiency as a function of n9 cut and detector size (the fiducial cut is applied). %
}%
\label{fig_Neutron_eff}
\end{figure}

\subsection{Detected signal and background events}

One source of IBD-like backgrounds are accidental coincidences of gamma-rays and neutrons generated by radioactive decays occurring in detector materials. %
These 
consist of two categories: 
\begin{itemize}
    \item Volume --- originate due to radioactive contaminants in the water (Rn/U/Th/K)
    \item Surface --- originate from the PMTs due to radioactivity of the glass
\end{itemize}
A full chain of simulations was performed to find the rates of these backgrounds for a representative set of detector sizes up to 50$\times$50.  
Secular equilibrium was assumed between the concentrations of U/Th and the resultant daughter decays.
In RAT-PAC, the full U/Th decay chains are included. 
The assumed radiopurities for water and PMT glass are listed in Table~\ref{tab_radiopurity}. The PMT values were obtained from measurements of Hamamatsu R7081 PMTs. The water values were chosen to be consistent with previous water-based experiments such as SNO~\cite{SNO2009} and Super-K~\cite{Abe:2016nxk, Ikeda:2019pcm}.

\begin{table}[h]
    \centering
    \begin{tabular}{|l||c|c|c|c|}\hline
Medium & $^{238}$U & $^{232}$Th &  $^{40}$K & $^{222}$Rn \\\hline\hline
Water [Bq/kg]	   & 1.0e--6	& 1.0e--7  & 4.0e--6    & 1.0e--6 	\\\hline
PMT [Bq/PMT]          & 2.45e3 	    & 2.49e3     & 5.85e--1  &  ---  	        \\\hline
   \end{tabular}
    \caption{Radiopurity levels used in the calculations.}
    \label{tab_radiopurity}
\end{table}

The detectors considered here were assumed to be placed at a depth roughly consistent with the proposed depth of the WATCHMAN experiment in the Boulby mine ($\sim$2.8 km water equivalent). At this depth, and assuming the detector fiducial volumes are protected by 4 meters of veto and PMT buffer as described above, 
cosmogenic fast-neutron backgrounds are expected to be subdominant. 
Muons when interacting with the detector medium 
can create 
long-lived isotopes (e.g. $^8$He and $^9$Li), that can potentially look like IBD events. 
At the Boulby depth, these can be effectively vetoed with minimal loss in livetime.

The rate of fast neutrons originating from muons interacting with the surrounding rock material was estimated using the Mei-Hime model~\cite{Mei:2005gm}. 
A FLUKA simulation was performed to determine 
the rate and positions of neutron captures inside the 20$\times$20 detector, and then scaled as the fiducial area. %
We estimate the rate of di-neutron events (that could mimic the IBD) as follows:
\begin{equation}
    M_i = \sum_j A_{ij} P_j
\end{equation}
\begin{equation}
    A_{ij} = \epsilon^i (1-\epsilon)^{j-i} 
    {j \choose i} 
\end{equation}
where $M_i$ --- measured multiplicity (to detect two neutrons, $i=2$), $P_j$ --- produced multiplicity (to produce $j$ neutrons), ${j \choose i} = \frac{j!}{i! (j-i)!}$ is the binomial coefficient, and $\epsilon$ is the detection efficiency taken from Fig.~\ref{fig_Neutron_eff} (a mean value is chosen corresponding to optimal n9 values for prompt and delayed).

The rates of exotic backgrounds such as from diffuse relic supernova antineutrinos and atmospheric neutrino-induced neutral-current interactions are unknown, though they are constrained to be  small~\cite{beacom_vagins_2004, Leyton:2018vff, PhysRevD.99.032005}. As such, these were estimated as follows: 4 events/10kt/yr (atmospheric neutrinos scatter off oxygen nuclei) and 2 events/10kt/yr (diffuse supernova relic antineutrinos).
The uncertainties on these two types of backgrounds are expected to be significantly smaller in the near future, as the SuperK-Gd will likely measure them within the next couple of years.

\begin{figure}[ht!]
\centering\includegraphics[width=1.\linewidth]{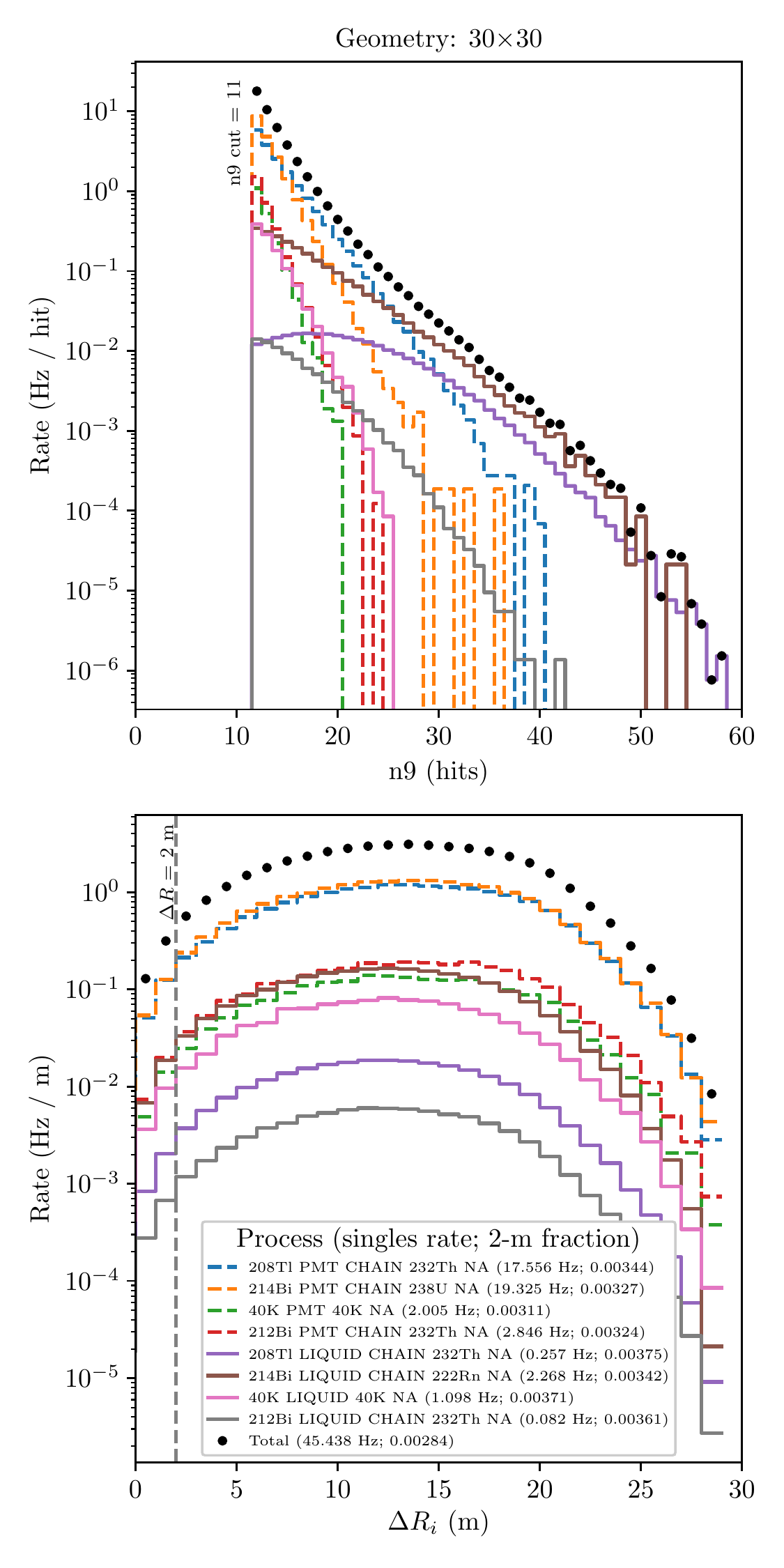}
\caption{Uncorrelated detector background (singles rate) inside the fiducial volume as a function of the n9 parameter for 30$\times$30 geometry. The primary contributing processes in the PMTs and water are shown. {\it Bottom panel:} Also for a 30$\times$30 detector, an example distribution (for an n9 cut of 11) of the distance between any two consecutive uncorrelated detector background events that reconstruct inside the fiducial volume. A line is drawn at the 2-m distance cut. The legend is shared between the two plots.
}
\label{fig_accidentals}
\end{figure}

\begin{figure}[ht!]
\centering\includegraphics[width=1.\linewidth]{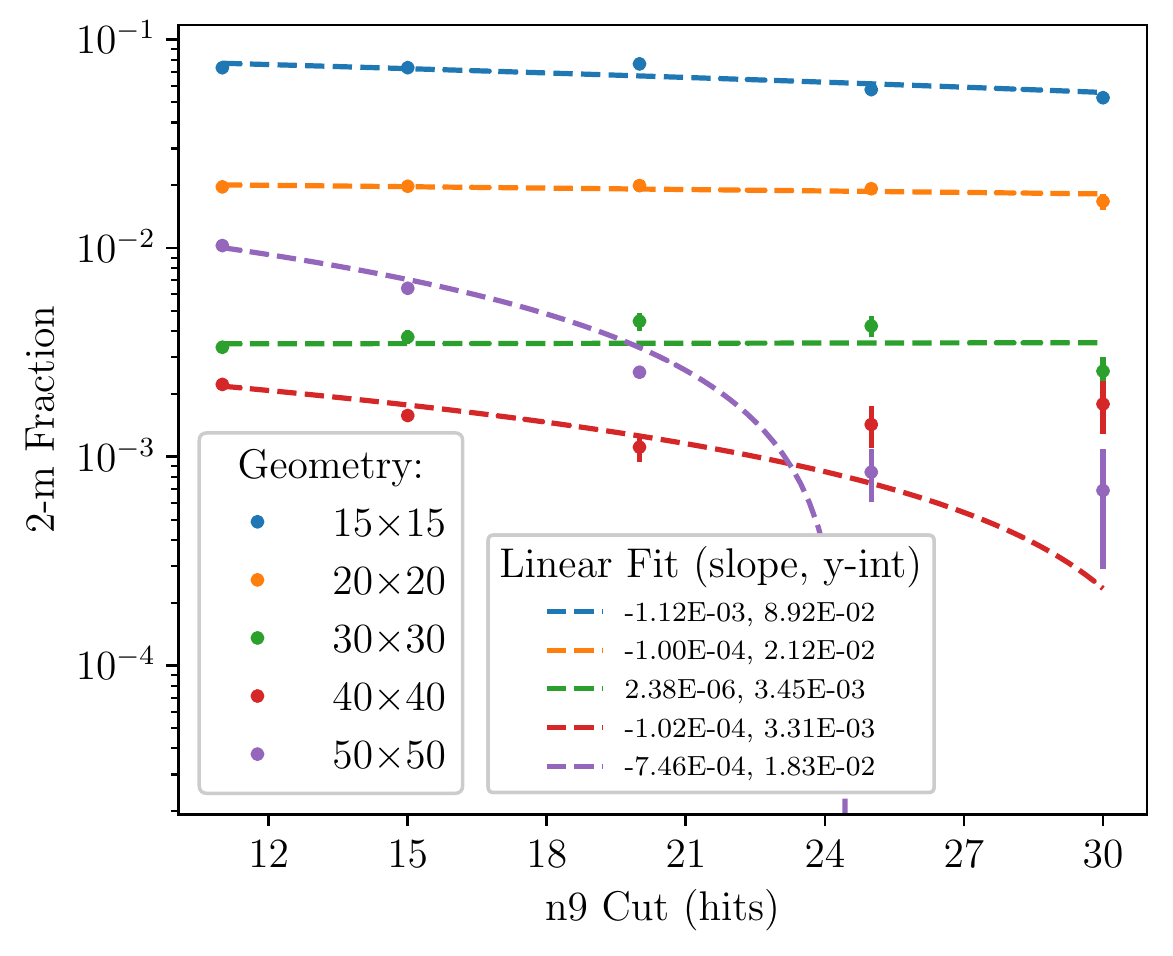}
\caption{Fraction of detector-induced accidental backgrounds events within a 2-m volume as a function of n9 cut for different geometries. Error bars are calculated as follows $\delta f = 1 / \sqrt{\sum N_i R_i / R_{tot}}$, where  $N_i$ is a number of events within a 2-meter distance of each other; $R_i$ --- rate of $i$-th process; $R_{tot}$ --- total rate (after applying fiducial-volume and n9 cuts). }
\label{fig_2m_bubble}
\end{figure}

\begin{figure}[ht!]
\begin{flushleft}
\includegraphics[width=0.49\linewidth]{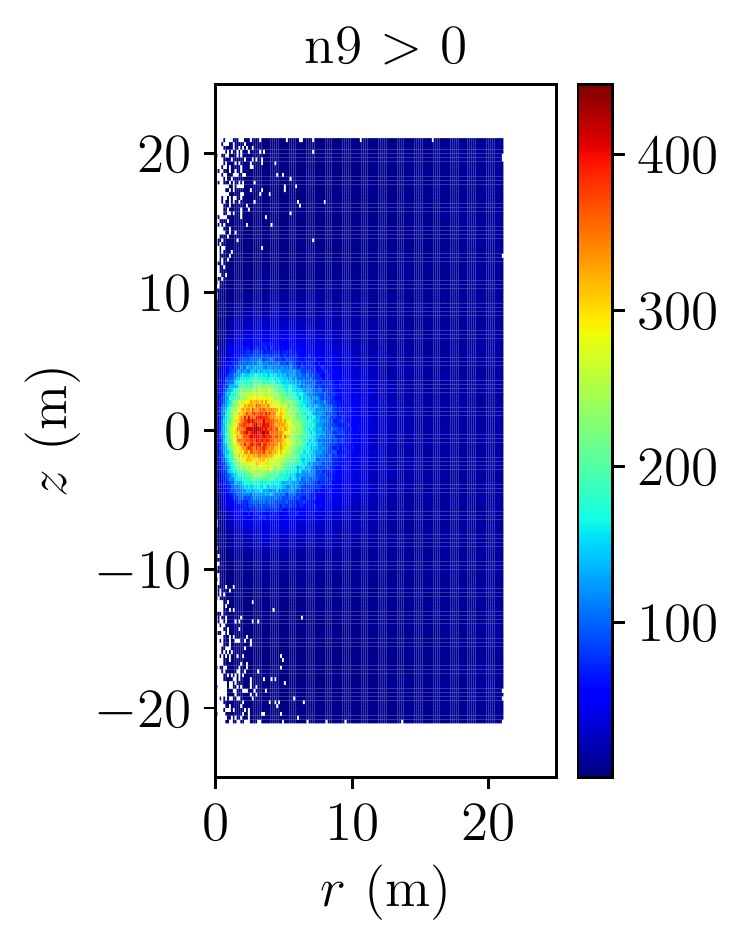}
\includegraphics[width=0.49\linewidth]{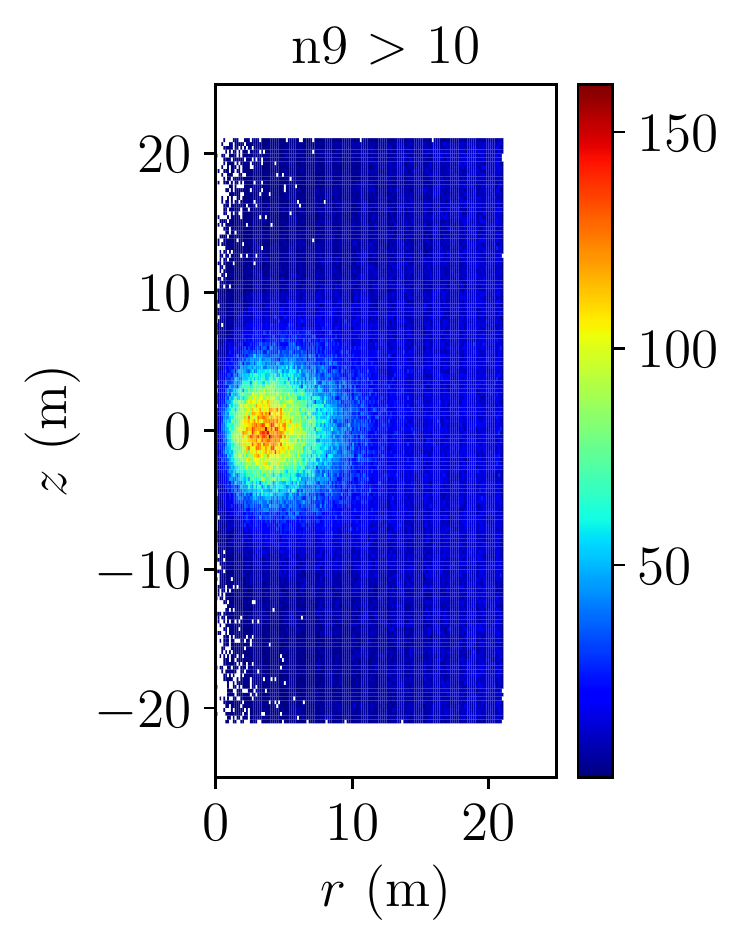}
\includegraphics[width=0.47\linewidth]{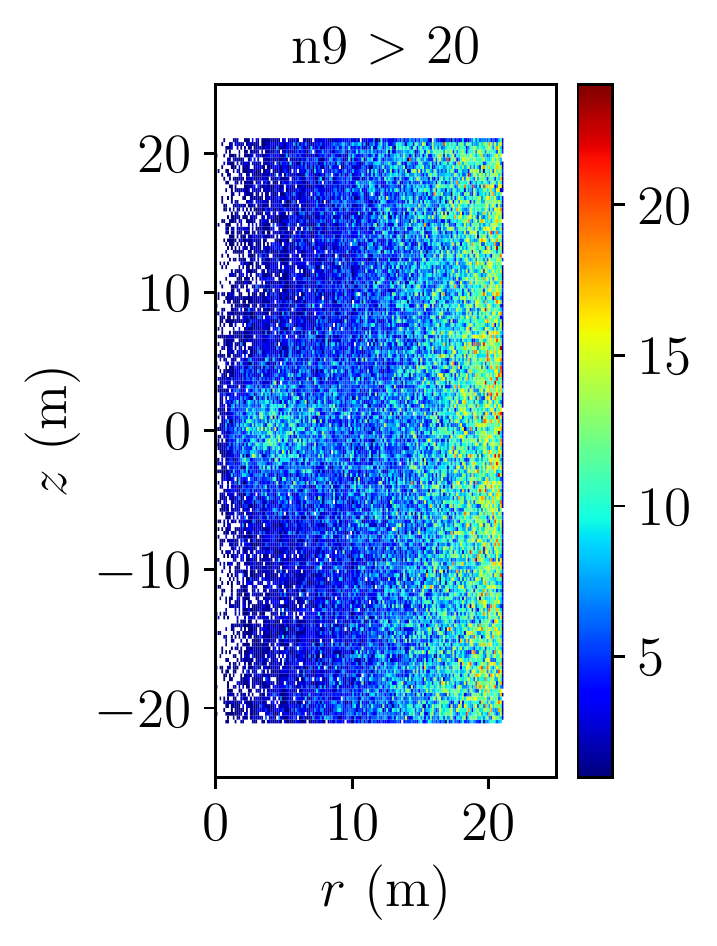}
\hspace{0.009\linewidth}
\includegraphics[width=0.47\linewidth]{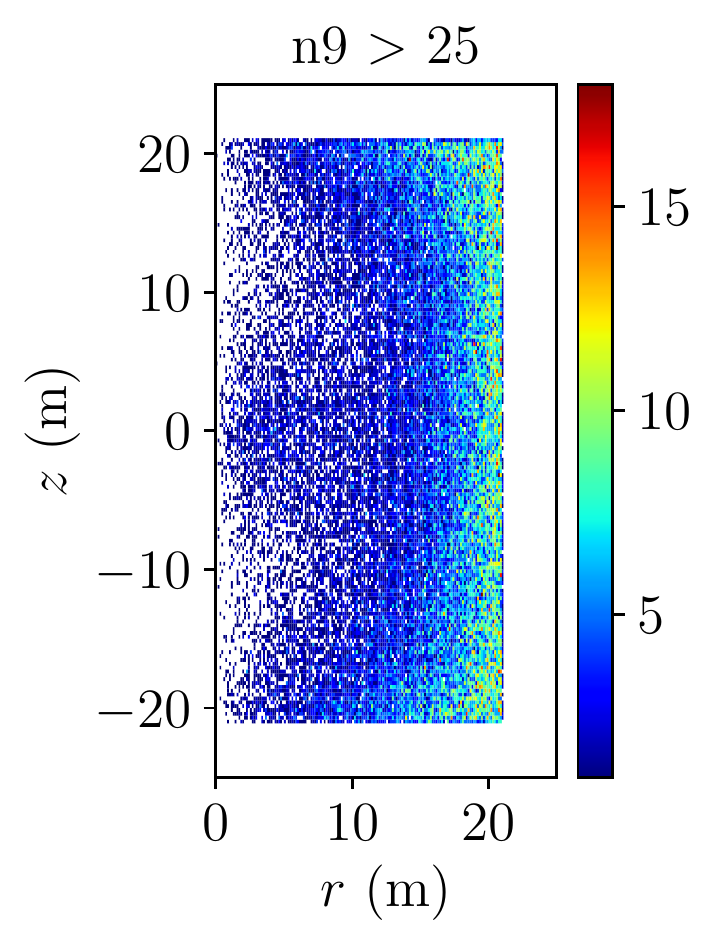}
\end{flushleft}
\caption{Spatial distributions of events originating in water due to $^{208}$Tl decays in the 50$\times$50 detector (reconstructed position $z$ vs $r=\sqrt{x^2 +y^2}$). The events that reconstruct in the middle of the detector ({\it top panel}) are due to the high PMT dark rate affecting the reconstruction algorithm. If the value for n9 cut is increased, the reconstruction works better.} %
\label{fig_darkrate_ZR_208Tl}
\end{figure}

In principle, the requirement that IBD event pairs be reconstructed within 2 meters places a significant limitation on the number of uncorrelated detector background events that result in an IBD-like event (see Fig.~\ref{fig_accidentals}). 
The fraction of events that pass this criterion was calculated from simulation for a selection of representative detector sizes and plotted as a function of n9 cut. The results are presented in Fig.~\ref{fig_2m_bubble}. The expectation was that for each detector size this fraction 
would be independent of the n9 cut. However,
for larger detector sizes 40$\times$40 and 50$\times$50, it appears that these two parameters are not independent. The reason appears to be for large detectors, there are a significant number of PMT hits caused by the PMT dark rate (see Figs.~\ref{fig_2m_bubble}~and~\ref{fig_darkrate_ZR_208Tl}). 
These dark-rate-induced PMT events severely impact event reconstruction, causing many low n9 events to be pushed towards the center of the detector. 
While dark rate appears to be the cause, the effect can be removed by applying a higher n9 analysis cut (see Table~\ref{tab_geom_optimal_param}).
For the 50$\times$50 geometry, an optimal n9 cut was chosen 26 to eliminate the effect of the PMT dark rate.
To determine the minimum n9 analysis cut, we calculate 
the ratio of the 2-meter volume divided by the fiducial volume. When the n9 cut applied in the simulation results in an equivalent 2-meter fraction (within uncertainty), it was assumed sufficient to correct for the effect.

For extremely large detector geometries, such as those  greater than 50$\times$50, 
the PMT dark rate of 3~kHz begins to impact the  vertex reconstruction algorithm. 
The effect could be seen in Fig.~\ref{fig_flat_positron_n9} and Fig.~\ref{fig_2m_bubble}.
For these detectors, either new ways must be found to reduce PMT dark rates, or larger-diameter PMTs should be used.
The high dark rate affect the reconstruction algorithm. For large detectors, events start to reconstruct at the center of the detector, as demonstrated in Fig.~\ref{fig_darkrate_ZR_208Tl}.

A scan over n9 values was performed to determine the optimal n9 threshold cut for our set of representative detector sizes (shown in Fig.~\ref{fig_accDwellTime}). The optimization parameter was the signal divided by the square root of the signal plus background. Only detector backgrounds were used since the cuts were chosen to optimally remove detector backgrounds while maximizing the reactor antineutrino signal. The optimal n9 thresholds for any general detector size were determined by interpolation from these results.

\section{Results and discussion}

\begin{figure}[ht]
\centering\includegraphics[width=1.\linewidth]{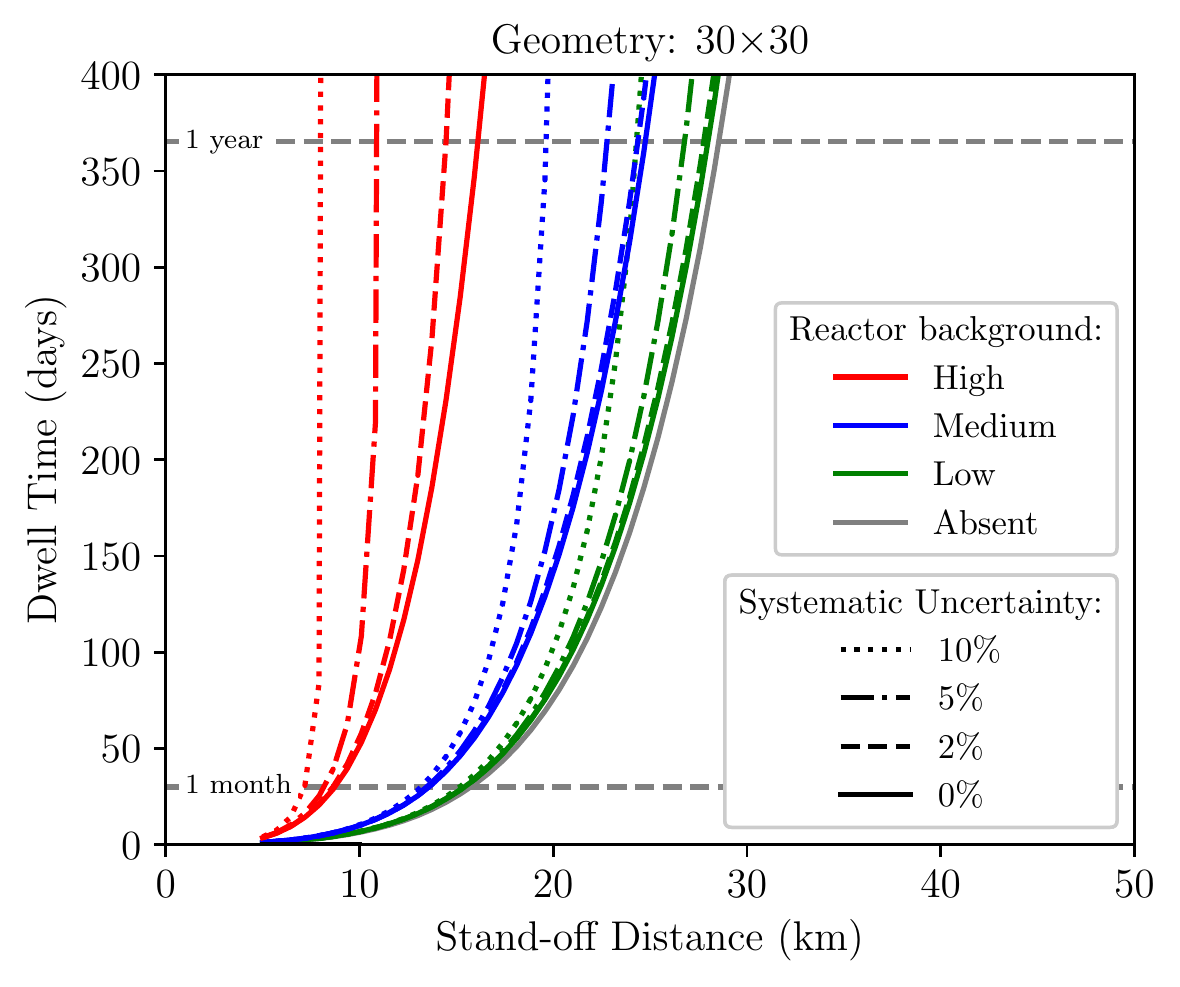}
\centering\includegraphics[width=1.\linewidth]{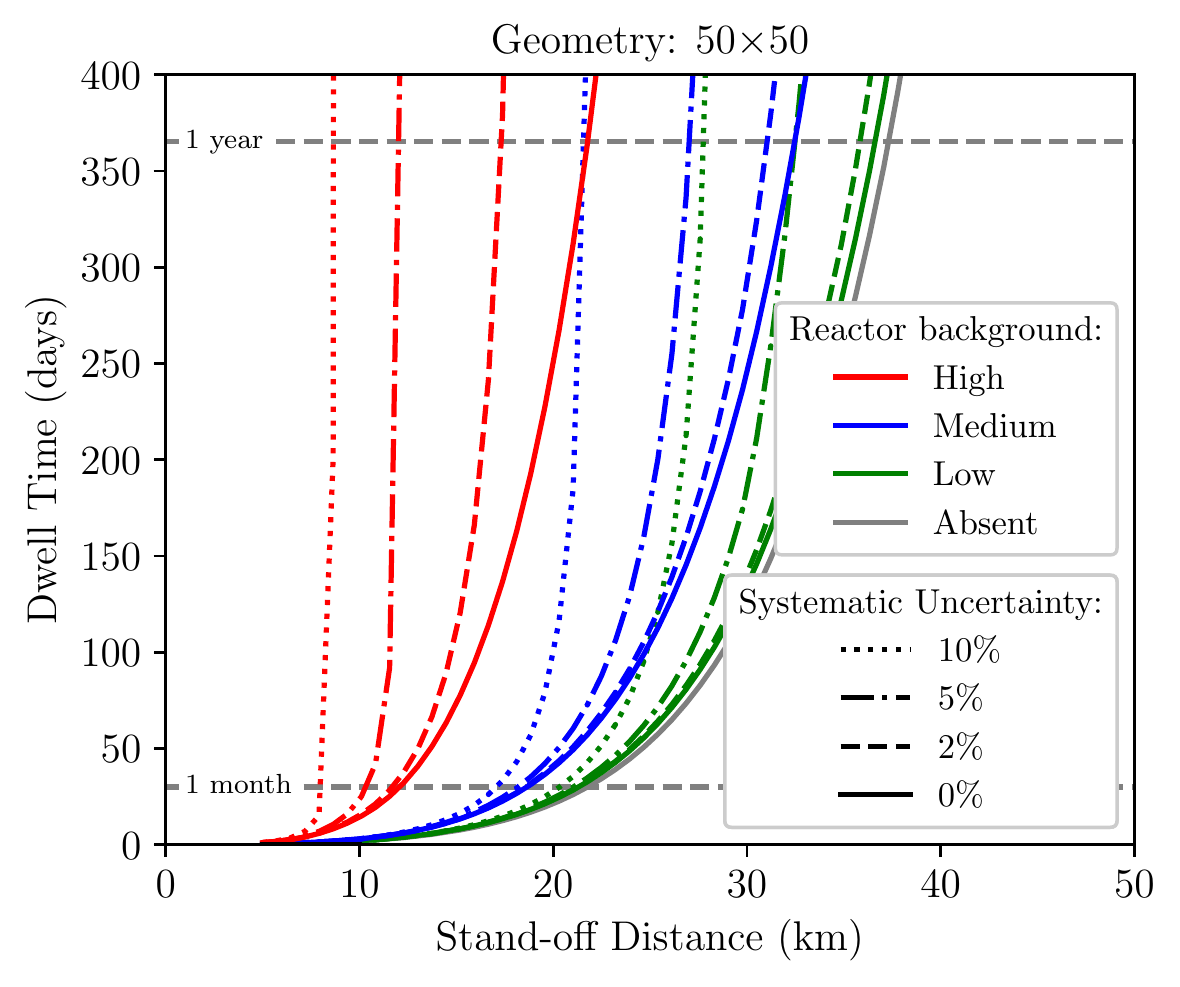}
\caption{Dwell time for two different geometries (30$\times$30 and 50$\times$50) as a function of stand-off distance to the 50-MWt reactor. The color represents world-reactor backgrounds. The line style represents four different levels of systematic uncertainties (0\%, 2\%, 5\%, and 10\%), based on Eq.~\ref{eq_currie_sys}. For clarity, the systematic effects are not shown for the ``absent'' world-reactor-background curve. 
The Gaussian assumption remains valid for dwell times greater than approximately one month. 
For smaller dwell times, a Poisson treatment is more accurate.}
\label{fig_dwellT_sys}
\end{figure}

We use the Currie equation~\cite{currie_eq, Knoll:2010xta} to evaluate the number of signal events required to register a significant detection.
The Currie equation can be generalized to the total uncertainty on the background, including the statistical and systematic uncertainty
as follows:
\begin{equation} \label{eq_currie}
    N_D = 4.653 \sigma_B  + 2.706
\end{equation}
The total uncertainty $\sigma_B$ has statistical and systematic uncertainties added in quadrature:
\begin{equation}
    \sigma_B^2 = \sigma_{stat}^2 + \sigma_{sys}^2 = N_B + (N_B \delta)^2
\end{equation}
and so,
\begin{equation} \label{eq_currie_sys}
    N_D = 4.653 \sqrt{N_B} \sqrt{1 + N_B \delta^2} + 2.706
\end{equation}
where $N_D$ is the minimum number of counts from the source (antineutrinos from a 50-MW reactor) required to ensure a reliable detection in the presence of background --- $N_B$ is the total background, including world reactors, uncorrelated detector backgrounds, cosmogenic fast neutrons, atmospheric neutrino interactions on oxygen, diffuse supernova antineutrinos and geological antineutrinos.
We assume that systematic uncertainty $\delta$ is symmetric, ensuring that the underlining assumptions used to derive the Currie equation remain valid (i.e. systematic and statistical uncertainties added in quadrature and the total systematic uncertainty remains Gaussian).
For low-count circumstances such as for short dwell times or 
long base-lines, systematic uncertainty $\delta$ can have an important impact on $N_D$.
The dwell time as a function of distance and the effect of adding non-zero systematic uncertainties ($\delta =$ 2\%, 5\%, and 10\%) are shown in Fig.~\ref{fig_dwellT_sys}.

\begin{table*}[ht!]
\centering
\begin{tabular}{|c|c||c|c|c||c|c|c||c|c|c|c||}
\hline
\multicolumn{2}{|c||}{Detector} & \multicolumn{3}{c||}{Signal (50-MWt)} & \multicolumn{7}{c||}{Backgrounds} \\\hline
\multirow{2}{*}{D$\times$H} & \multirow{2}{*}{$m_{\mathrm{fid}}$/kt}&  \multirow{2}{*}{10-km} & \multirow{2}{*}{20-km} & \multirow{2}{*}{50-km} &  \multicolumn{3}{c||}{World-Reactor} & \multirow{2}{*}{Accidental} & \multirow{2}{*}{Cosmogenic} & \multirow{2}{*}{Exotic} & \multirow{2}{*}{Geo} \\\cline{6-8}
 & & 
& & &
Low & Med. & High &
& & & 
\\\hline
\hline
20$\times$20 &1.3 &  
89 & 20 & 2  &
2   & 19   & 248   &  
2   & 24  & 1 & $<$1 \\\hline
30$\times$30 &8.4 &  
498 & 113 & 9  & 
11  & 105  & 1388  &  
7  & 81  & 5  & 2 \\\hline
40$\times$40 &25.7 &  
1232 & 280 & 23 &
26  & 258  & 3397  &
17  & 165 & 15 & 3 \\\hline
50$\times$50 &58.2 &  
1614 & 372 & 34 & 
33  & 324  & 4017  &
2   & 247 & 35 & 3 \\\hline
\hline
\end{tabular}
\caption{Number of counts per year for signal and various backgrounds in the simulated detector geometries. The signal annual rates are reported for 10, 20, and 50-km baselines.} 
\label{tab_counts}
\end{table*}

\begin{figure}[ht]
\centering\includegraphics[width=1.\linewidth]{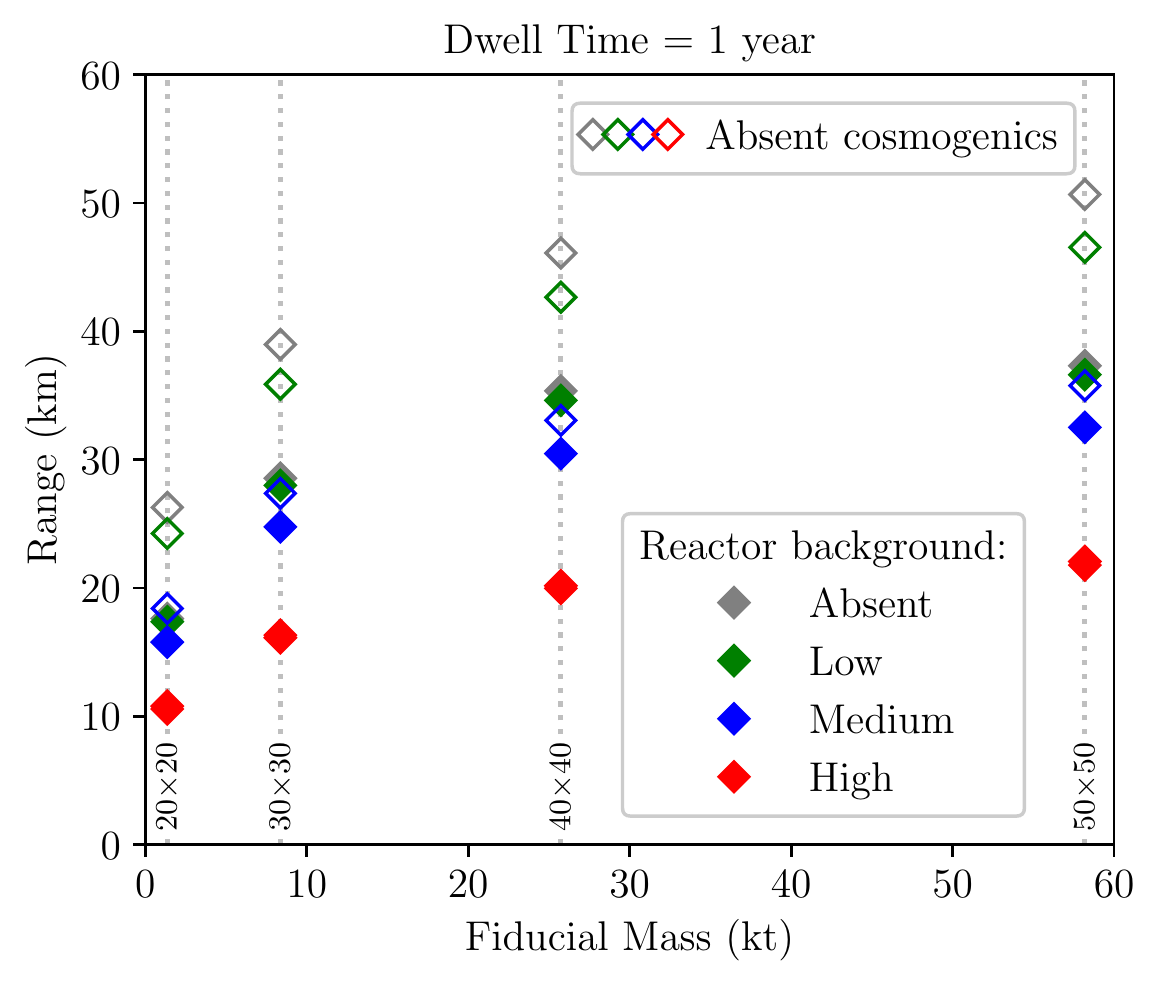}
\caption{Range as a function of stand-off distance for a 1-year dwell time at the three world-reactor backgrounds. The empty rhombi indicate scenarios with zero cosmogenic fast-neutron backgrounds.}
\label{fig_dwellT_1year_range}
\end{figure}

The flux and spectrum of reactor antineutrinos for any reactor--detector distance was calculated by applying the up-to-date neutrino-oscillation parameters as shown in Fig.~\ref{fig_osc}.
Since the world-wide reactor flux background can vary considerably depending on where in the world the detector is placed, three locations were chosen as broadly representative of three reactor background cases: high, medium and low. %
The three locations chosen were Andes, Baksan, and Frejus underground laboratories. The web tool, geoneutrinos.org, was used to calculate the world-reactor antineutrino fluxes in those locations~\cite{Barna:2015rza}. %
Unlike detector backgrounds, antineutrino backgrounds cannot be removed by applying clever analysis cuts, and so the world's reactors comprise an important and irreducible background in these calculations. %

Other important performance parameters needed for this analysis were the detector background rates as a function of n9 (energy) (Fig.~\ref{fig_accidentals}), and the fraction of uncorrelated detector backgrounds that reconstruct inside the fiducial volume and within the 2-m proximity requirement for IBD events. Fig.~\ref{fig_accidentals} also shows the distribution of uncorrelated detector backgrounds as a function of the distance from previous event for a variety of different detector backgrounds. 

For high world-reactor background locations (e.g. the Frejus site at the border between France and Italy), the task of detecting a small 50-MWt reactor at a large distance
depends primarily on the world-reactor background at that location. %
This is the primary reason why the estimated range levels off as a function of detector size 
in Fig.~\ref{fig_dwellT_1year_range}.

All other backgrounds, such as those associated with detector radiopurity and cosmogenic interactions were found to be subdominant for the Frejus location. 
Although geological antineutrinos were included in the calculations, they were easily rejected, as their energies were either below the detection threshold, or not far enough above it to produce an energetic enough event for detection.
Table~\ref{tab_counts} shows the number of signal and background events for each detector location.
The signal values presented in Table~\ref{tab_counts} were calculated using  Eq.~\ref{eq_IBD_rates}, where the positron, neutron, spatial, and temporal efficiencies were based on Fig.~\ref{fig_nuebar_spectrum_3locations}, Fig.~\ref{fig_Neutron_eff}, and Table~\ref{tab_geom_optimal_param}.

One significant outcome of this work is that it is clear that water-Cherenkov detectors, despite the clarity of the detection medium, seem to approach a sensitivity limit at the 50$\times$50 scale (as shown in Fig.~\ref{fig_dwellT_1year_range}). Detectors at this scale appear to level off in sensitivity at $\sim$50 km due to a combination of neutrino oscillations, which become significant at that distance, PMT dark rate, and detector self absorption, which appears to impact performance at the 50$\times$50 scale. The problem of the negative effect of PMT dark rate on event reconstruction severely impacts sensitivity in this work. However, improvements in event reconstruction algorithms might be able to overcome this difficulty~\cite{Kneale:2021}. Such improvements, if achievable, were considered outside the scope of this work.  

\section{Acknowledgements}
This work is supported by the U.S. Department of Energy National Nuclear Security Administration and Lawrence Livermore National Laboratory [Contract No. DE-AC52-07NA27344, release number LLNL-JRNL-823255].
We thank the WATCHMAN collaboration for useful comments and critique at various stages of this study. 
In particular, we thank Felicia Sutanto for performing the FLUKA fast-neutron simulation.
V.L. thanks Mark Duvall for the help with visualization of large geometries in RAT-PAC.
All computations were performed on the Livermore-Computing Lassen cluster running Red Hat Linux operating system.
Analysis routines were written using ROOT~6~\cite{Brun:1997pa} and Python 3, primarily with NumPy~\cite{numpy2020}, SciPy~\cite{Virtanen:2019joe}, Pandas~\cite{pandas2010}, and Matplotlib~\cite{Hunter:2007} packages.

\nocite{*}
\bibliography{refs_href}

\end{document}